\documentclass[prb,twocolumn]{revtex4}
\usepackage{graphicx,amsmath}

\begin{document}
\title{Uncertainty relations  on the slit and Fisher information }
\author{Z. Hradil}
\author{J. \v{R}eh\'{a}\v{c}ek}
\affiliation{Department of Optics, Palacky University, 772~00 Olomouc,
Czech Republic}
\date{\today}

\begin{abstract}
Diffraction on the slit can be interpreted in accordance with
the Heisenberg uncertainty principle. This elementary example
hints at the importance of the information theory for the quantum physics. 
The role played by one particularly interesting measure of information ---
the  Fisher information --- in quantum measurements is further discussed in 
the context of quantum interferometry.
\end{abstract}
\maketitle

Quantum mechanics became a standard tool not only of physicists,
but almost any scientist is familiar with at least some of its
concepts. Traditionally it has  provoked and attracted attention
of broader community due to sometimes paradoxical implications on
the fundamental and philosophical issues of the Nature. That is
why teaching of quantum mechanics  is a rewarding but not an easy
task. There is perhaps no other field of physics encumbered by so
many misconceptions  and misinterpretations as quantum theory is.
Every student realizes very quickly that depending on the degree of
profundity there are no simple questions and answers here.
Recently there was a live discussion concerning the teaching of
quantum mechanics \cite{Muller-Wiesner2002} followed by a comment
\cite{Hilgerwood2002} addressing a misunderstanding 
appearing in many
elementary textbooks of quantum theory, namely the exposition of
the Heisenberg uncertainty relation on the slit. We do not want to
speculate whether it is better on the introductory level  to
suppress the complexity of the problem and present a simple but
physically inadequate explanation, or to  built a  consistent
but terse theory from the very beginning. This is certainly a deep
problem, solution of which depends on the teacher. The purpose of
this contribution is however different. The analysis of the
diffraction of the wave (particle) on the slit
will be used as a link between the quantum mechanics and mathematical
statistics. We  will  provide a physical motivation for the
concept of  the Fisher information and will reconsider the
problem of uncertainty relations related to the quantum interpretation
of the interference pattern in wave theory. In our opinion, both
these issues deserve attention: Fisher information comes from
the mathematical statistics and as such it seems to do very little
with quantum theory. That is perhaps why this topic is missing in standard
textbooks. This may change in the future 
since now the crucial role of information
physics is widely recognized, and there exist serious attempts
to derive and explain  the physical laws of nature and quantum theory itself
from the theory of information \cite{Frieden,caslav1,caslav2}. 

The second goal  of our approach is to formulate
alternative inequalities for complementary observables 
motivated by the Fisher information 
and apply them to particles passing through the slit. 
This intuitively clear and
illustrative example will demonstrate  the incompatibility of
the position and impulse observables in quantum theory. It is intriguing
to note that this approach is also relevant for such up to date
problems as quantum interferometry and quantum lithography 
\cite{yablonovitch,Kok00,Kok01}.

\section{Diffraction on the slit}

\begin{center}
\begin{figure}
\centerline{\includegraphics[width=0.9\columnwidth]{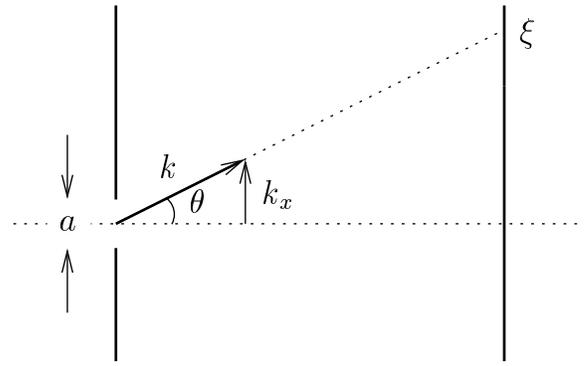}}
\caption{Geometry of the diffraction on the slit.\label{fig:slit}}
\end{figure}
\end{center}

Let us consider the standard setup and standard argumentation used
in the elementary textbooks of quantum mechanics for the 
exposition of Heisenberg uncertainty relations \cite{Greiner}. For simplicity
assume 1D geometry of a single slit sketched in Fig.~\ref{fig:slit}. 
The particle goes through the slit impinging on the position sensitive
screen behind the slit. According to the de Brogli hypothesis
it will effectively behave as a wave with the de
Brogli wave length $ \lambda = h/ p$, $p $ being the impulse of
the particle. Using a simple geometrical argumentation, each detection
event on the screen may be used for inferring the direction of
the incoming particle. Hence this scheme could be considered as a
measuring device for determining the impulse.
 Invoking the effect of diffraction,
the quantum nature of particles will be manifested by a diffraction
pattern registered on the screen. This effect will enhance the uncertainty of
the inferred impulse.  This qualitative reasoning can easily be
accompanied by the corresponding wave picture that describes 
each detected single event. Assuming the illumination by a plane wave,
the state describing the particle behind the slit reads
\begin{equation}\label{state}
  \psi(x) = \left\{
  \begin{array}{ll}
  \exp(i k_x x)/\sqrt{a}, \quad &|x|\le a/2,\\[2mm]
   0,\quad & |x|> a/2. 
   \end{array}
   \right.
\end{equation}
Here $k_x = k \sin\theta$ is the component  of the wave vector 
$ k = 2\pi/\lambda$ orthogonal to the optical axes, and $ \lambda $ 
is the wavelength of the particle.  
Denoting the detected  position of
the particle on the screen by  $\xi$, the probability of
the detection of the particle in the far reach zone is given by
the square of its  Fourier transformed wave function,
\begin{equation}\label{probability}
  p(\mu) = \frac{1}{\pi} {\rm sinc}^2(\mu - \nu).
\end{equation}
Here the dimensionless quantities used are  $\mu = \frac{a k }{2
d} \xi$ and $\nu = a/2 k_x$. Naturally,  such a
detection visualizes the transversal momentum of the particle $k_x$  
impinging on the slit.  The probability manifests one distinct peak at
$\mu=\nu$. The standard   interpretation relies on a geometrical
argumentation. Taking the first minimum of the function 
\eqref{probability} for defining its ``spatial extent,'' 
the half-width of the probability distribution is then determined as 
$ \Delta k_x = \frac{2 \pi}{a} $. Considering further that in the plane
of the slit the  location of the particle is known to be within the
half-width of $\Delta x = a/2$, the expected uncertainty relation
reads
\begin{equation}\label{uncert-approx}
  \Delta p_x \Delta x \approx h/2,
\end{equation}
where $p = \hbar k$. This is often considered as a painless, quick
and intuitive way how to formulate the  uncertainty relation,
more so, because the relation (\ref{uncert-approx}) resembles the famous
uncertainty relation of quantum theory ascribed to Heisenberg
\begin{equation}\label{uncert}
  \Delta p_x \Delta x \ge \hbar/2,
\end{equation}
though its exact derivation in the present form has been done by
Kennard \cite{Kennard}.  One must however realize that 
contrary to the relation
(\ref{uncert-approx}), the uncertainties in (\ref{uncert}) are
strictly defined  as the root-mean-square variances of observables.
Due to the formal similarity between both the relations one may be
tempted to interpret the  relation (\ref{uncert-approx}) as an
approximation of the rigorous uncertainty principle (\ref{uncert}).
 Unfortunately, this would not be correct, since, 
 the whole derivation of \eqref{uncert-approx} stands on the shaky ground. 
 The error measures appearing it
(\ref{uncert-approx}) have nothing to do with standard variances.
This becomes crucial in the case of the probability distribution
(\ref{probability}) whose variance is infinite due to its heavy tails.
This problem is well known \cite{Nussenzweig}, and usually  
ascribed to the discontinuity of the wave function 
after the passage of the particle through the slit, and
in fact, it disqualifies the example of the slit from all exact
considerations. Obviously, as the uncertainty of the momentum
is infinitely large, inequality (\ref{uncert}) is trivially satisfied, 
in this case. There are several ways how to circumvent this obstacle,
for example by alternative definitions of the proper resolution
measure, or by invoking entropic uncertainty relations
\cite{entropic_unc}.

In the following exposition we will keep the example of the diffraction on
a slit as a toy example and employ it for introducing the Fisher
information. This information defines the ultimate limitations of 
measurements, and as such it can be used to describe the uncertainty
relations in the generalized sense. As an interesting byproduct,
the resolution of the current quantum interferometric techniques will be
evaluated.

\section{Fisher information}

The interference pattern registered  behind the slit may be
interpreted within a different framework. The build-up
of an interference pattern is governed by a probabilistic law,
where the intensity (\ref{probability}) plays the role of
a probability distribution. Instead of a single-particle detection
discussed in the previous section, let us consider the information
accumulated in  the detection of the  full interferometric pattern
created by altogether $n$ particles. It does not matter whether they
arrived in a single shot or one by one in the course
of a subsequently repeated experiment. The identity of detected
particles is disregarded and this is the only additional
assumption with respect to the previous case. Let us develop
the statistical description  capable of handling a generic statistical
model, which will afterwards be applied to the problem of the 
diffraction on the slit.

Assume that generic data denoted by  $x$  are registered with 
the frequency of occurrence of $n_x$, and let us denote the
total number of particles by $n$. For the concreteness,
the variable $x$ is considered to take discrete values. 
Moreover, we assume that the values $x$ occur with  
the probability $ p_x(\bar \theta), $ where $\bar \theta$ represents 
an unknown true value of a certain parameter. The purpose of the whole
treatment is to infer the true value of this parameter as faithfully as
possible from the registered data $x$. This is the general
estimation scheme, which of course, might be adopted to the case
of particles impinging on a screen as well. In particular,
provided that the detection of each single particle is
evaluated separately, $n=1$, this scheme reduces to the above mentioned 
measurement of the impulse. Let us denote by $\theta$  a function of 
the registered data, which will be used for the estimation of the 
true value of the parameter of interest $\bar \theta.$ This function is 
called estimator in the mathematical statistics and there are many
nonequivalent ways how to construct it. Let us describe the
maximum likelihood (ML) estimator adopted to the evaluation of the
the accumulated data set ${n_x}$. Since registrations of individual
particles are independent events, the
likelihood that the actual value of the parameter was
$\theta$ conditioned upon the registered data is proportional to the
product of individual probabilities,
\begin{equation}
\label{likeli}
 {\cal P}(\theta | \{ n_x \} ) \propto 
 \exp \Bigl\{\sum _x  n_x \ln p_x(\theta)\Bigr\}.
\end{equation}
ML estimator is given by such a value of $\theta$ which
maximizes  this function. Let us estimate its uncertainty.
Provided that the total number $n$ of registered particles is
large, the registered data
can be replaced by the expected number of detected particles, $ n_x = n
p_x(\bar \theta)$. The likelihood \eqref{likeli} can then be expanded
in a power series in the neighborhood of this true value,
\begin{eqnarray}\label{distribution}
  {\cal P} &\propto& \exp\Bigr\{ n \sum_x p_x(\bar \theta) \ln p_x(
  \theta)\Bigr\} \\ 
  &\approx& \exp\biggl\{ n \sum_x p_x(\bar \theta) \ln
  p_x(\bar \theta)\nonumber \\ 
  &&- \frac{n}{2}  \sum_x \frac{ p'^2_x( \bar
  \theta)}{p_x(\bar \theta)} ( \theta -\bar \theta )^2 + \ldots
  \biggr\}.\nonumber
\end{eqnarray}
Its meaning is obvious. For sufficiently large number of
particles the ML estimator fluctuates around the true value of
the parameter within the error $ 1/F$,
\begin{equation}
   F_{\mu}= \sum_x \frac{ p'^2_x( \bar \theta)}{p_x(\bar \theta)}.
\end{equation}
$F$ is the Fisher information, which characterizes the root mean
square error of the inferred value of the parameter from its true
value. Significantly, the Fisher information represents the
ultimate limit for the resolution of any unbiased estimator.
This relation is known as the Cramer-Rao lower bound \cite{rao,cramer}.
For its importance we repeat its derivation here \cite{helstrom}.
Noticing that the mean value of any unbiased estimator equals the true 
value,
\begin{equation}\label{unbiased}
\sum_x\theta \, p_x(\bar\theta)=\bar\theta,
\end{equation}
and using the normalization condition 
\begin{equation}\label{normalized}
\sum_x  p_x(\bar\theta)=1,
\end{equation}
this inequality can be derived by differentiating Eqs.~\eqref{unbiased}
and \eqref{normalized} with respect to $\bar\theta$,
multiplying the latter result by $\bar\theta$,
and subtracting it from the former, 
which gives
\begin{equation}\label{before_cs}
\sum_x(\theta-\bar\theta)p'_x=1.
\end{equation}
The expression on the left-hand side is bounded from above by 
the Cauchy-Schwarz inequality,
\begin{equation}\label{after_cs}
1\le\Bigl[\sum_x(\theta-\bar\theta)^2 p_x\Bigr]\Bigl[ 
\sum_x\frac{(p_x')^2}{p_x}\Bigr]=(\Delta\theta)^2 F,
\end{equation}
so finally, we get
\begin{equation} \label{CR}
(\Delta \theta)^2 \ge  1/F,
\end{equation}
which is the Cramer-Rao lower bound on the variance of an unbiased estimator.
Of course, to achieve this resolution it is necessary to register
a large number of particles $n$. As follows from the expansion of
the likelihood, the performance of a ML estimator 
improves with increasing number of particles as $1/(nF).$ 
Hence the Fisher information gives the
ultimate resolution corresponding to a single "average" particle
from the bunch of registered events \cite{nas}.

\section{Fisher information of interference patterns}

Let us apply  the  theory to the interference pattern behind the
slit. It is easy task to calculate all the respective quantities
\begin{eqnarray}
 (\Delta \mu)^2 = (\frac{a}{2\hbar})^2 (\Delta p_x)^2,
 \label{unc1}\\  F =
\frac{4}{\pi} \int d\mu ( \frac{d}{d \mu } {\rm sinc}\, \mu)^2 =
\frac{4}{3},\label{unc2}\\
(\Delta x )^2 = \frac{a^2}{12}.\label{unc3}
\end{eqnarray}
What is really intriguing, the Cramer-Rao inequality (\ref{CR})
reproduces exactly the expected Heisenberg uncertainty relations
for impulse and position of the particle going through the slit.
Indeed, plugging the uncertainties \eqref{unc1}, \eqref{unc2},
and \eqref{unc3} into Eq.~\eqref{CR} the resulting 
uncertainty relation reads
\begin{equation}\label{Heisenberg}
  \Delta p_x \Delta x \ge \frac{\hbar}{2}.
\end{equation}
Remarkably, the state of particle behind the slit meets the
equality sign here being the minimum uncertainty state. The price
paid for this interpretation is the reinterpretation  of the
measurement of the impulse.  On the contrary to the single
detection case, here the identity of separate particles is
disregarded. What we observe is the impulse of the "center of mass" of
the bunch of particles rather than the impulse of each particle
separately.

But there are still some other remarkable  differences in
interpretation. Provided that accuracy is related to $n$-particle
signal (for example, assuming $n$-particle absorption process), the
accuracy improves according to the  distribution
(\ref{distribution}) $n$ times and the same effect appears in the
Cramer-Rao inequalities (\ref{CR}). According to the standard
interpretation, this improvement is viewed as the effect of
a rescaling of the de Brogli wavelength due to the $n$ times greater
mass of interfering  ``quasi'' particles \cite{yablonovitch,Kok00,Kok01}.  
The improved accuracy is
consequently referred to as a measurement beyond the Heisenberg limit,
which of  course seems to be problematic in the view of the above
mentioned arguments.

Another point is also intriguing:  The standard exposition
of the Heisenberg uncertainty relations relies on the notion of
measurable quantities. Students are taught that such observables
correspond to  hermitian operators and that the 
accuracy of such observations can be described by variances. 
Since both the variances of position and impulse appear in the standard
Heisenberg uncertainty relations, it may  seem at the first glance
that that uncertainty relations limit the simultaneous
(inaccurate) measurement of non-commuting observables. This is 
certainly not true. Heisenberg uncertainty relations express
merely a necessary condition obeyed by any wave function.

The problem of simultaneous measurement of non-commuting
observables is a more involved problem and is therefore  beyond the
exposition in the undergraduate course. It was first discussed
by Arthurs and Kelly \cite{Arthurs-Kelly} for position and impulse
observables, and the answer relies upon the notion of 
generalized measurements.
They showed that in the case of a simultaneous 
measurement of position and momentum, the product of the
uncertainties becomes two times larger
compared to the standard Heisenberg uncertainty relation.
It is interesting
to note that there is no danger of similar misinterpretation
within the information theory. Particularly, the  Cramer-Rao
inequality for the estimation  of a single variable (\ref{CR}) involves
a single  variance only. Hence the formulation anticipates the
measurement of single parameter only (impulse). The variance of
the position was  introduced in order to establish
the link to the standard Heisenberg uncertainty principle. Notice,
however, that Cramer-Rao inequalities may be easily extended to
higher dimensions, and particularly, the Cramer-Rao inequality for
simultaneous detection of impulse and position will reveal two
times higher right-hand side \cite{Arthurs-Kelly}.

There is a simple relationship between the Fisher information and
variances of complementary variables.  Therefore,  the Cramer-Rao
inequalities imply the  standard uncertainty relation \cite{Hall}.
Considering the momentum representation,
\begin{eqnarray}
\frac{\partial}{\partial p}\psi(p) = \langle p |
\frac{\hat X}{i \hbar}| \psi\rangle, \\
\frac{\partial}{\partial p}\psi(p)^* =  - \langle \psi  |
\frac{\hat X}{i \hbar}| p\rangle,\\
\psi(p) = \langle p | \psi\rangle, \quad \psi(p)^* = \langle \psi | p\rangle,
\end{eqnarray}
the Fisher information may be rewritten to the form
\begin{eqnarray}
F_{p} = \int d \mu \frac{[\psi(\mu)^{ *\prime} \psi(\mu) +
\psi(\mu)^{\prime} \psi(\mu )^*]^2}{ \psi(\mu) \psi^{*}(\mu)} \\
= \int d{\mu} \frac{1}{\psi(\mu) \psi^{*}(\mu)}
\bigl\langle \mu \bigl| \bigl[ \frac{\Delta  \hat X}{i \hbar}, 
|\psi \rangle \langle \psi | \bigr]\bigr|
\mu \bigr\rangle^2.
\end{eqnarray}
Using the identity 
$$ |\langle \hat A \hat B \rangle |^2 = \frac{1}{4}
|\langle [\hat A, \hat B] \rangle |^2  + \frac{1}{4} |\langle
\{\hat A, \hat B\} \rangle |^2 ,  $$
we get
\begin{eqnarray}
F_{p} &=& \frac{4}{\hbar^2} \langle \psi |(\Delta X)^2| \psi
\rangle
\\
\nonumber  &-&     \int \!d \mu\,
 \psi(\mu) \psi^{*}(\mu) \biggl[ \frac{\partial }{\partial \mu}
 \arg(\psi(\mu)) + \frac{1}{\hbar}  \bar X\biggr]^2,
\end{eqnarray}
where $ \bar X = \langle \psi| \hat X |\psi
 \rangle  $. That is why the  Cramer-Rao inequality is 
 stronger
than the Heisenberg uncertainty relation,  since always
\begin{equation}\label{stronger}
 (\Delta p)^2 \ge \frac{1}{F_{p}} \ge 
 \frac{\hbar^2}{4(\Delta x)^2}. 
\end{equation}
They will coincide whenever the phase of the wave function is
related to the mean value of position by the relation
\begin{equation}\label{phase}
\frac{\partial }{\partial \mu}
 \arg(\psi(\mu)) = - \frac{1}{\hbar} \bar X,
\end{equation}
which means that the phase of the wave function of minimum uncertainty states in
$p$-representation exhibits a linear dependence on the
momentum, $\arg \psi(\mu) = \alpha + \beta \mu$,  and, in particular,
it must be constant if $ \bar X = 0 $. This condition 
gives a wide class of minimum uncertainty states for 
the informatic uncertainty relation on the slit.

\section{Quantum interferometry}

\begin{center}
\begin{figure}
\centerline{\includegraphics[width=0.9\columnwidth]{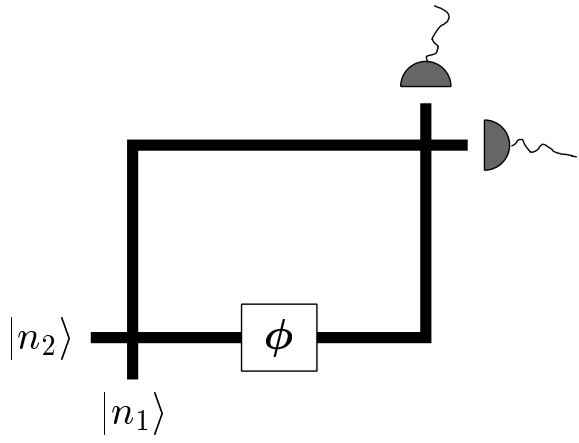}}
\caption{Mach-Zehnder interferometer.\label{fig:mz}}
\end{figure}
\end{center}

The analysis of the diffraction on the slit from the information
point of view may serve not merely as
an alternative interpretation of this experiment 
but it could also be used for  the evaluation of the performance
of interferometric techniques.
The variance of the detected signal is crucial
for the maximal attainable resolution.
This can be seen from the following simple linearized
theory, frequently used in various considerations. Let us
assume the measurement of a generic  operator $A $.
This measurement should reveal the value of an unknown parameter
$\bar \theta$ appearing in a unitary  transformation $U(\bar
\theta) $ applied to a quantum state $|\psi\rangle$: 
$|\psi(\bar \theta) \rangle = U(\bar
\theta) | \psi \rangle.$  Provided that the average value of the
$\bar \theta $- dependent signal $\langle A(\bar \theta) \rangle $
is detected, the desired value of the parameter may be inferred by
means of an estimator $\theta$. The estimated value is usually uncertain
since the signal itself fluctuates with the variance given
by $(\Delta A)^2 = (\langle A^2 \rangle - \langle A \rangle ^2 $.
Symbol  $(\Delta)^2 $ will be reserved for variances, 
whereas symbol $\delta $ will denote an
error obtained by means of a linearized theory,
\begin{equation}\label{lin}
\delta \theta   \approx  \frac{{\Delta A}}{ \langle A(\bar\theta)\rangle'
},
\end{equation}
where prime denotes a derivative with respect to the parameter.  
For its simplicity, the latter measure of error is frequently used in 
interferometry \cite{Yurke86}.
However, one should keep in mind that in some cases the two
quantities  can be significantly different. 
Let us evaluate a simple realistic
model of the phase detection by means  of a Mach-Zehnder interferometer,
see Fig.~\ref{fig:mz}.
Formally, this device can be described by Lie algebra SU(2), the 
correspondence  being provided by the Schwinger representation of the 
angular momentum operators, 
\begin{eqnarray}
\label{angular} J_1 = \frac{1}{2}(a_1^{\dagger} a_2 +a_2^{\dagger}
a_1), \\ \nonumber  \quad J_2 = \frac{1}{2i}(a_1^{\dagger} a_2 -
a_2^{\dagger} a_1), \\ \nonumber  J_3 = \frac{1}{2}(a_1^{\dagger}
a_1 - a_2^{\dagger} a_2).
\end{eqnarray}
All these operators commute with the
total number of particles $ N = a_1^{\dagger} a_1 + a_2^{\dagger}
a_2$ and satisfy commutation relations of angular momentum observables 
$ [J_i, J_j] = i \epsilon_{ijk} J_k. $ Provided that a difference of
particle numbers on the output ports will be registered, the
measurement is represented  by $ J_3$ operator. Before reaching
the detectors, the input signal undergoes subsequent
transformations on two beam-splitters and phase shifter.
Considering a symmetrical beam splitter and merger, both introduce 
the  phase shift of $\pi/2$ for the reflected signal, 
their action on the input state of light is described by the unitary
operator $ e^{-i \pi J_1 }.$ 
Phase shifter transforms the state according to the
unitary transformation $e^{-i(\phi_2 -\phi_1) J_3 }.$ 
Consequently,
the quantum state at the output carries the information about
phase shift $\phi = \phi_2 -\phi_1 .$ The total transformation
induced by the Mach-Zehnder interferometer then reads,
\begin{eqnarray}
\label{U-transf}
 U(\phi) = e^{i \pi/2 J_1} e^{-i \phi J_3} e^{- i
\pi/2 J_1} = e^{-i \phi J_2},\\
\label{state2}
|{\mathrm out}\rangle  =  U(\phi)|{\mathrm in}\rangle,\\
\label{transform}
U^{\dagger}(\phi) J_3 U(\phi) = -\sin \phi  J_1 + \cos\phi J_3.
\end{eqnarray}

Assume now that the input ports are fed by   $n_1$ and $ n_2$  particles,
respectively. Adopting  the standard  notation of eigenstates
$|j,m \rangle $ of the operators $J^2$ and $J_3$, 
the input state simply  reads $ |\mathrm{in}
 \rangle =|j, m\rangle $, where  $j = (n_1 +n_2) /2$ and $m= (n_1
-n_2)/2 $.
The output of the measurement can either be characterized by 
phase dependent moments
\begin{eqnarray}
\label{J-3}
  \langle J_3\rangle =
m \cos \phi,  \\
\label{J^2}
\langle J_3^2  \rangle = m^2 \cos^2\phi + 1/2
[j(j+1) -m^2] \sin^2\phi,
\end{eqnarray}
or, more completely, by phase dependent  output statistics, 
\begin{equation}\label{sampled}
p_k(\phi) = |\langle j,k |U(\phi) |j,m  \rangle |^2,
\end{equation}
sampled by the measurement.
Such a measurement is obviously complete $\sum_k p_k(\phi) =1 $.
In the following, the description will further be simplified
considering only phase shifts  near the working point $\phi=0$,
where the ultimate limit of the accuracy of the phase-shift measurement 
can be evaluated most easily.
The Fisher information associated with the phase estimation
in this case reads,
\begin{equation}\label{Fisher-phase}
  F_{0}  = \sum_{k} \frac{[p_k^{\prime}(0)]^2}{p_k(0)},
\end{equation}
Notice that all probabilities $p_k(\phi) = | \langle j,k| e^{-i \phi J_2}
| j, m \rangle |^2$, evaluated at $\phi = 0  $ vanish
except when $k= m$. 
In the former case, this leads to an indefinite expression
of the type $0/0 $ under the summation sign in Eq.~(\ref{Fisher-phase}).
Evaluating them with the help of the 
L'Hospital rule we obtain,
\begin{equation}\label{Fisher-phase2}
  F_{0}  = \sum_{k\neq m} \frac{2 p_k^{\prime} p_k^{\prime \prime}}{p_k^{\prime}}
  = 2 \sum_{k\neq m} p_k^{\prime \prime}
  = - 2 p_m^{\prime \prime} ,
\end{equation}
since  $\sum p_k =1.$ Consequently, for the input state 
$|j,m\rangle $ and true phase $\phi =0 $, we found the Fisher 
information to be
\begin{equation}\label{F_0}
  F_0 = 4  \langle j,m | J_2^2| j,m\rangle  = 2 [j(j+1) - m^2] .
\end{equation}
At the same time the phase error could be 
estimated by means of the simple linearized theory as
\begin{equation}\label{phase_error}
(\delta \phi)^2  = \frac{(\Delta J_3)^2 }{(J_3(\phi)^{\prime})^2 }
=  \frac{j(j+1) - m^2}{2 m^2}.
\end{equation}
Now assume two cases of special interest
related  to the  classical and quantum regimes.
Provided that the interferometer is
operated in the usual (classical) manner with the light entering one 
input port only, $|{\mathrm in} =\rangle |j=\frac{N}{2}, m= \frac{N}{2} \rangle
$, the  phase error  given by the linearized theory reads 
$\delta \phi = 1/\sqrt{N}$. 
Significantly, the ultimate variance predicted by the Cramer-Rao
inequalities gives the same value of
$\Delta \phi = 1/F_0 = 1/\sqrt{N}.$
This regime is usually referred to as the standard limit of phase 
measurements.

Intriguing situation appears when both the  input ports of the
interferometer are  fed by  the signal with an equal number of
particles $N/2$. In this case we have state 
$|j=\frac{N}{2}, m= 0 \rangle $ at the input 
and the situation becomes  diametrically different. 
Here, the simple linearized theory fails to provide an error estimate
because $\delta \phi = 0/0 $.
However, the information approach can still be used and the 
Fisher information predicts the ultimate phase resolution
of 
\begin{equation}\label{quantum}
 \Delta \phi = \sqrt{2}/N.
\end{equation}
This regime is referred to as  the quantum limit of phase measurements
\cite{sanders}

Let us clarify some experimental aspects of this quantum resolution regime
\cite{myska}.
The probability distribution of the detected signal can be approximated for 
large values of $j$ by
\begin{equation}\label{bessel}
\mathcal{P}(\phi|m ) = p_m(\phi) \approx  J_m^2(\phi j),
\end{equation}
where $J_m$ denotes the  Bessel function. According to the Bayes
principle  this expression also provides the  posterior phase 
distribution conditioned on the detected photon number difference $m$. 
Obviously, the most accurate detection of the phase shift is
expected for a carefully balanced interferometer ($\phi =0$) when input
state is transmitted without any change $m=0$. However, an
experimentalist is facing an inverse problem. Provided that  the value
$m=0$ was registered, one cannot be sure whether the true value of the phase 
shift had really been set to zero. Of course, $m=0$ could be detected
with some probability for other phase shifts as well. 
There is always some uncertainty about the estimated parameter.
As a proper measure of performance, the width of the respective posterior phase 
distribution can be adopted. Accepting the same heuristic argumentation 
as in the case of the diffraction on the slit, the first minimum of the
Bessel function could be used to define the width. Indeed, the
resolution obtained in this way is of  the order of
$(\delta \phi)^2 \approx 1/ j^2;$ $N= 2 j$ which resembles the quantum 
phase resolution beating  the standard noise limit of phase \cite{HB93}. 
However, neither in the case of diffraction, nor 
here the distance from the center to the first minimum of the probability 
distribution is a plausible measure of error in quantum theory.
One should realize that distribution (\ref{bessel}) is not even square 
integrable  in the limit of $j \rightarrow \infty. $ Its 
variance can be evaluated  assuming the phase window $\pi$:
\begin{equation}\label{p-variance }
  (\Delta \phi)^2  =
   (\frac{1}{j})^2 \frac {\int_{0}^{j\pi/2} dx J^2_0(x) x^2}
{\int_{0}^{j\pi/2} dx J^2_0(x)} \propto 1/\ln j.
\end{equation}
Here the Bessel function was approximated by its asymptotic
expansion in the last step, since the heavy tails of the distribution
yield a dominant contribution. What does this mathematical
expression mean? Provided that the true phase is estimated from 
the result of a single shot measurement, the result is rather
uncertain \cite{phase_analys}. 
In particular, such measurement is much worse compared 
to the interferometer  operated in the usual manner 
(with  single illuminated input port).  
However, our scheme still posses some potential for an improvement. 
Let us imagine that our single-shot measurement is repeated. 
Provided the most optimistic result $m=0$ is detected again, the
posterior distribution will now become
\begin{equation}\label{bessel2}
P(\phi|m=0;m=0 ) \approx J_0^4(\phi j),
\end{equation}
where $\{m=0; m=0 \}$ represents accumulated phase-sensitive data.  
The procedure of data accumulation can be
repeated until the optimal resolution is reached \cite{myska}.  As can be
shown, the accuracy predicted by the Fisher information  is achieved 
provided that the single shot detections are repeated $n$ times, $n \ge 4.$
For instance, in the most optimistic case when 
$\{m=0; m=0;m=0;\ldots m=0 \}_n $ was detected,
the variance of the posterior phase distribution is $ (\Delta \phi
)^2  =  1/ (n j^2). $ Since the total number of particles
used in the experiment is $ N_{tot}= N n$, the phase  resolution 
scales with the total energy used as
\begin{equation}\label{quant-res}
(\Delta\phi )^2 = 4n/N_{tot}^2, \quad n\ge 4.
\end{equation}
As can be seen the optimum regime is that with $n=4$ repetitions. 
Notice however, that although the quantum phase resolution $\propto 1/N^2$  
has been recovered, the optimal variance
is still by a constant factor larger than the ultimate limit \eqref{quantum}
predicted by the Fisher information. Indeed, the data accumulation 
that is needed to find the tiny sharp central peak of the phase
distribution always tends to move the resolution from the quantum toward the 
classical regime; in the extreme case of $n=N_{tot}/2$
when the particles are injected into the interferometer in pairs
we only get the classical resolution of $(\Delta\phi )^2=2/N_{tot}$. 
The fact that the ultimate Fisher limit cannot be attained 
is in accordance with the theory, since
this ultimate limit need not be achieved by measurement.
 This analysis illustrates the application of the  concept of Fisher information
in practice.

The setup of quantum interference similar to quantum
interferometry with equal number of particles beams  has recently been
proposed for  beating the Rayleigh resolution limit in lithography
\cite{Kok00}. The goal of the quantum lithography is
to overcome the well-known limitation of classical optics and 
create a pattern in one and two dimensions whose resolution
is below the diffraction limit.
As any finite aperture cuts off the spatial frequencies higher
than the inverse Rayleigh distance, the information about small
details beyond this diffraction limit is inevitably
lost. This may appear as detrimental, for instance, in
nanotechnology.  The proposal of quantum lithography is motivated
by the same setup used in quantum interferometry. As has been
demonstrated above, the quantum interference pattern can be
created by means of non-classical interference.
 The phase shift is induced  by the
differences of optical phases inside the interferometer. The
argument  of Bessel function depends on the energy of  photon
number correlated beams. The shape of pattern    can be compressed
beyond the standard diffraction limit, provided that the substrate
is sensitive to N-photon absorption. This technological obstacle
makes  this technique rather tricky and  speculative, at least, from the
technological point of view. However, the  argumentation based on
Fisher information raises other questions.

The width of the distribution in the sense of the distance to the
first minimum (Rayleigh criterion) can certainly be decreased. However, as
the price, only a small portion of the whole probability (or energy)
concentrates there. The remaining
portion of the signal is spread out and contributes to the
overall noise. This undesirable effect will obviously influence
the possibility to shape the desired pattern due to the
unavoidable fluctuations. For  applications in lithography one
needs the possibility to create and transmit high
frequencies as well as the capability to point to a target.

\section{Conclusions}

We have formulated several arguments in favor of Fisher
information and its applications to quantum problems.
Fisher information provides the ultimate limitation for quantum measurements
and as such also provides a nontrivial link between 
the theory of statistics and quantum theory.
Since quantum theory is more ''operational'' than perhaps any other
physical theory, this may yield new interesting insights
into its fundamentals.
From the pedagogical point of view, it is especially interesting to 
apply the Fisher information to simple thought experiments
frequently discussed in elementary textbooks on quantum mechanics. 
Particularly, we have shown that the example of the
diffraction on the slit can be interpreted as the  uncertainty
principle for the impulse and position of particles creating the
diffraction pattern. Similar argumentation can be used for
the description of the phase resolution of quantum interferometry. 

\acknowledgments{
This work has been supported by projects LN00A015  and J14/98
of the Czech Ministry of Education, by the EU grant QIPC, Project No.
IST-1999-13071 (QUICOV), and by Czech-Italian project No.~29,
``Decoherence and Quantum Measurement.''.}

\newpage

\vspace{8cm}

\vspace{5cm}

\begin{verbatim} RehacekFig1.eps \end{verbatim}

\newpage

\vspace{8cm}

\vspace{5cm}

\begin{verbatim} RehacekFig2.eps \end{verbatim}
\end{document}